\documentclass[structabstract]{aa}  
\usepackage{txfonts}
\usepackage[dvips]{graphicx}
\usepackage{natbib}
\usepackage{longtable}

\def\fermi{\textit{Fermi~}}
\def\fermilat{\textit{Fermi}/LAT~}

\bibpunct{(}{)}{;}{a}{}{,}
\begin{document}

   \title{Correlation between \fermilat gamma-ray and 37 GHz radio properties of northern AGN averaged over 11 months}

 \author{E. Nieppola \inst{1,2}
          \and
           M. Tornikoski \inst{2}
          \and
           E. Valtaoja \inst{3}
          \and
           J. Le\'on-Tavares \inst{2}
          \and
           T. Hovatta \inst{4}
          \and
           A. L\"ahteenm\"aki \inst{2}
          \and
           J. Tammi \inst{2}
          }

   \institute{
Finnish Centre of Astronomy with ESO (FINCA), University of Turku, V\"ais\"al\"antie 20, FI--21500 Piikki\"o, Finland \\ \email{elina.nieppola@aalto.fi}
	\and
     Aalto University Mets\"ahovi Radio Observatory, 
       Mets\"ahovintie 114, FI--02540 Kylm\"al\"a, Finland
         \and
    Tuorla Observatory, Department of Physics and Astronomy, University of Turku, FI--20100 Turku, 
     Finland
          \and
     Department of Physics, Purdue University, 525 Northwestern Ave., West Lafayette,
     IN 47907, USA
             }

   \date{Received ; accepted }

  \abstract{}{Although the \fermi mission has increased our knowledge of gamma-ray AGN, many questions remain, such as the site of gamma-ray production, the emission mechanism, and the factors that govern the strength of the emission. Using data from a high radio band, 37 GHz, uncontaminated by other radiation components besides the jet emission, we study these questions with averaged flux densities over the the first year of \fermi operations.}
{We look for possible correlations between the 100 MeV -- 100 GeV band used by the \fermi satellite and 37 GHz radio band observed at the Aalto University Mets\"ahovi Radio Telescope, as well as for differences between the gamma-ray emission of different AGN subsamples. We use data averaged over the 1FGL period. Our sample includes 249 northern AGN, including a complete sample of 68 northern AGN with a measured average flux density exceeding 1 Jy.}
{We find significant correlation between both the flux densities and luminosities in gamma and radio bands. The \fermi luminosity is inversely correlated with the peak frequency of the synchrotron component of the AGN spectral energy distributions. We also calculate the gamma dominances, defined as the ratio between the gamma and radio flux densities, and find an indication that high-energy blazars are more gamma-dominated than low-energy blazars. After studying the distributions of gamma and radio luminosities, it is clear that BL Lacertae objects are different from quasars, with significantly lower luminosities. It is unclear whether this is an intrinsic difference, an effect of variable relativistic boosting across the synchrotron peak frequency range, or the result of \fermi being more sensitive to hard spectrum sources like BL Lacertae objects. Our results suggest that the gamma radiation is produced co-spatially with the 37 GHz emission, i.e., in the jet.}{}


	\titlerunning{Correlations between the \fermi and 37 GHz bands}

   \maketitle

%

\section{Introduction}

Only 10\% of active galactic nuclei (AGN) exhibit significant emission in radio wavelengths. Non-thermal radio emission is not an intrinsic feature of the nucleus itself, but it always indicates the presence of relativistic jets. These jets emanate from the centre symmetrically, often stretching kiloparsecs into intergalactic space. By harbouring a population of relativistic electrons, the jets produce broad-band synchrotron emission from the radio to X-ray frequencies that swamps any thermal radiation from the galaxy. Another dominant radiation component in the spectral energy distribution (SED) of blazars is inverse Compton (IC) radiation. The IC photons have energies between X-ray and TeV regions. They are produced through the interaction of the relativistic electrons, which also produce the synchrotron component, and a population of seed photons. The origin of these seed photons is unclear. One possible scenario is synchrotron self-Compton (SSC) radiation, where the seed photons are the electrons' own synchrotron photons. The other option is external inverse Compton (EC) radiation, for which the seed photons come from a source external to the jet. This source could be the broad line region (BLR) \citep{sikora94}, accretion disk \citep{dermer93}, or the molecular torus surrounding the nucleus \citep{blazejowski00, sokolov05}. Thus, typically the EC scenario requires that the IC radiation, i.e., the gamma-rays, originate relatively close to the centre, within the central parsec, while the SSC gamma-rays come from the jet beyond a parsec's distance. The radio emission, however, is undoubtedly produced in the jet. If we can establish a firm correlation between the radio and gamma-ray emission of AGN, that would provide support that they are likely to have a co-spatial origin, and the gamma-rays are produced most likely through the SSC mechanism. In turn, the lack of correlation would indicate that the gamma-rays are produced close to the nucleus independently of the synchrotron photons, and the EC scenario would be a stronger candidate. There are, however, theories including EC photon sources well away from the central engine, such as the cosmic microwave background radiation \citep{tavecchio00}, possible outflowing BLR \citep{leontavares10}, and the tentative sheath of the jet \citep{jorstad10}. The IC radiation is probably always a mixture of the SSC and EC processes in some measure, but the SSC mechanism is certainly more likely downstream in the jet.

Until recently, the best gamma-ray data on offer was the one provided by the Energetic Gamma Ray Experiment Telescope (EGRET) aboard the Compton Gamma-Ray Observatory, operational between 1991 and 2000. Relying on these data, \citet{valtaoja95} showed that the production of gamma rays was indeed connected to the rise in the 37 GHz radio flux. That would indicate that emission in both frequency bands originates in the same area and that the most likely radiation mechanism for the gamma-rays would be SSC. This conclusion was supported by several later studies \citep{valtaoja96_gamma, jorstad01, lahteenmaki03}. 

However, the sensitivity and sampling of EGRET was not sufficient for a detailed study of the correlations. With the launch of the \fermi Gamma-ray Space Telescope in 2008 the situation has changed drastically. The primary instrument of \fermi is the Large Area Telescope, LAT. It is an imaging, wide field-of-view telescope covering the energies between 20 Mev and 300 GeV. \fermi works in survey mode, scanning the full sky in only three hours. It also has unprecedented sensitivity, the detection limit being $F(E>100\,\mathrm{MeV)} \simeq 7.5 \times 10^{-9}\,\mathrm{ph\,cm^{-2}\,s^{-1}}$ with high galactic latitude and photon index $\Gamma=2.2$, after 11 months of operation \citep{lott10}. \fermi finally offers the flux and time sensitivity needed to shed more light on the correlation between the frequency bands. Evidence that the \fermilat fluxes lead the radio fluxes has already been found by \citet{kovalev09}, \citet{pushkarev10}, and \citet{agudo11}. However, research supporting the EC scenario has also been published \citep{tavecchio10}. The Mets\"ahovi team has also revisited the results obtained in \citet{valtaoja96_gamma} and \citet{lahteenmaki03} using \fermi flux curves \citep{leontavares11}. The conclusion that at least the strongest gamma-rays are produced in the jet on parsec scales still stands with this improved data set. 

In this paper we connect \fermi data with 37 GHz radio data from Mets\"ahovi Radio Observatory. High radio frequencies are ideal for use in correlation studies, as they are produced co-spatially with the assumed SSC radiation. Lower frequencies ($< 15$ GHz) can be contaminated by the extended emission from the radio lobes, and separating the jet emission from the emission of the nucleus is a problem in the optical bands. Thus, we expect our results to give an accurate picture of the possible connections between the frequency domains. We study properties averaged over the 11-month period of the first \fermi catalogue, 1FGL \citep{abdo10_1fgl}. In addition to 37 GHz fluxes, we study the connection between the gamma fluxes and the synchrotron peak frequencies of the sample AGN. Individual pointings and flux curves are studied in \citet{leontavares11}.

\section{Sample}
\label{sam}

In our sample we included all AGN from the Mets\"ahovi source list that had been detected by \fermilat during the 1FGL period. Most sources are northern, 1730-130 being the most southern. The whole sample comprises 249 AGN. We have divided our sample further into different AGN classes to allow their comparison. We have 146 BL Lacertae objects (BLO), 38 quasars (QSO), 34 high polarization quasars (HPQ), 11 low polarization quasars (LPQ), 8 non-quasar radio galaxies (GAL), and 12 unclassified radio sources. The dividing line in the optical polarization between high and low polarized quasars has been taken to be 3~\%. For sources classified as quasars (QSO) we have not found any optical polarization measurement in the literature. 

To compensate for the stochastic nature of our sample, we also performed the analyses for a complete subsample. This sample includes all northern AGN whose average flux density since the beginning of the monitoring is above 1 Jy at 37 GHz. The length of monitoring varies: it can be three decades for the best-known sources, while for the newer sources in the sample just a few years. The complete northern sample consists of 68 sources (30 HPQs, 17 BLOs, 8 LPQs, 6 GALs, and 6 QSOs). Most of the analyses in this paper were conducted separately for the whole sample and the complete northern sample. We have listed our sample sources in Table~\ref{sample}. Columns 1 and 2 give the source name and its counterpart in the 1FGL catalogue, respectively. Column~3 lists the classification and the objects belonging in the complete sample are marked in Col.~4. The source coordinates are given in Cols.~5 and 6, and redshift in Col.~7.
\addtocounter{table}{1}

\section{Data}

Our radio data include unpublished 37 GHz flux density measurements from the Aalto University Mets\"ahovi Radio Observatory. Mets\"ahovi radio telescope is a radome-enclosed antenna with a diameter of 13.7 metres. It is situated in Kirkkonummi, Finland, at 60 m above sea level. The 37 GHz receiver is a dual-horn, Dicke-switched receiver with an HEMT preamplifier, and it is operated at ambient temperature. A typical integration time for obtaining one flux density data point is 1200--1600 seconds, and the detection limit under optimal weather conditions is about 0.2 Jy. For more details about the Mets\"ahovi observing system and data reduction, see \citet{terasranta98}. 

The gamma-ray data are taken from the \fermilat First Source Catalog \citep[1FGL,][]{abdo10_1fgl}. The flux densities in this study have been calculated with the formula \begin{equation}\label{flux}F_\gamma=\frac{S(E_1,E_2)}{(1+z)^{2-\Gamma}}\end{equation} as in \citet{ghisellini09}, and luminosities in the usual way \begin{equation}L_\gamma=4\pi d_L^2 F_\gamma.\end{equation} In the equations $S(E_1,E_2)$ is the gamma-ray energy flux between energies $E_1=100\,\mathrm{MeV}$ and $E_2=100\,\mathrm{GeV}$, $z$ the redshift, $\Gamma$ the photon spectral index, and $d_L$ the luminosity distance. All quantities, except for the luminosity distance, are listed in the 1FGL catalogue. The lack of redshift information meant we could only calculate the gamma-ray flux and luminosity for 190 sources. In the luminosity calculations we used the cosmology $H_0=70\, \mathrm{km\,s^{-1}\,Mpc^{-1}}$, $\Omega_m=0.27$ and $\Omega_\Lambda=0.73$. 

Because the data in the 1FGL catalogue were taken during the first 11 months of the \fermi mission from August 4 2008 to July 4 2009, we only included 37 GHz data from the same period. One hundred fourteen sources had relatively strong flux levels and were always detected ($S/N\geq 4$) during that period. In addition, 53 sources had at least one $S/N<4$ non-detection. For these sources we calculated upper-limit, average 37 GHz fluxes, which are clearly indicated in the figures and analyses when used.

\section{Results}
\label{results}

\subsection{Flux densities and luminosities}
\label{correlations}

\begin{figure}[]
\includegraphics[width=0.5\textwidth]{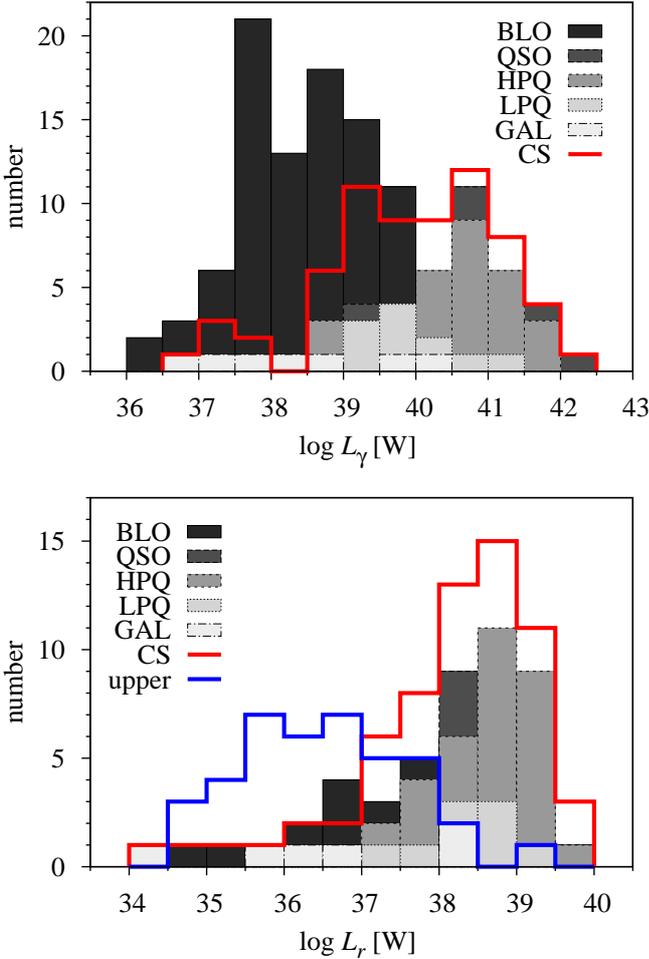}
\caption{Distribution of the gamma-ray (top panel) and 37 GHz radio (bottom panel) luminosities in the different subsamples. 3C 274 at log\,$L_\gamma=34.8 \mathrm{W}$ has been omitted from the top panel. CS in the legend stands for the complete northern sample.}
\label{radgamlum_hist}
\end{figure}

Figure~\ref{radgamlum_hist} shows the distribution of the \fermilat $\gamma$-ray and 37 GHz luminosities. The subsamples are plotted with different shades to illustrate their differences. The luminosity distributions of the complete northern sample are also plotted separately with a red line (abbreviated as CS in the legend). It is clear that BLOs have lower gamma-ray luminosities than quasars. The few radio galaxies in the sample are quite evenly distributed, although 3C 274 (M 87) at log\,$L_\gamma=34.8$\,W was omitted from the figure for clarity. Our complete northern sample is emphasized in the high-luminosity end. According to the Kruskal-Wallis test (performed with the Unistat statistical package v5.6.01), BLOs and galaxies are indeed drawn from a different gamma-ray luminosity distribution than all quasar subsamples. 

The radio luminosity histogram shows very similar characteristics. Here the number of BLOs is much lower because of their typically faint and often undetected radio fluxes at 37 GHz. That is why we chose to plot also the distribution of the upper limit radio luminosities, calculated for sources with $S/N\leq 4$ non-detections during the 1FGL period (blue line in the figure). Most of these (31/40) are BLOs. When these upper limit luminosities are taken into consideration, we see that the distribution of the radio luminosities is very similar to that of the gamma-ray luminosities. No clear differences between the 37 GHz luminosity distributions of the subsamples are found with the Kruskal-Wallis test. However, including the upper limit data points in the analysis changes the result. We performed two sample tests for censored data with the ASURV Rev 1.3 software \citep{lavalley92}, which implements the methods for censored data presented in \citet{isobe86}. According to logrank and Gehan's generalized Wilcoxon tests, the final result is the same as with gamma luminosities, and BLOs differ from all quasars groups at the 5\% level and share a similar parent distribution with galaxies.

\begin{figure}[]
\includegraphics[width=0.5\textwidth]{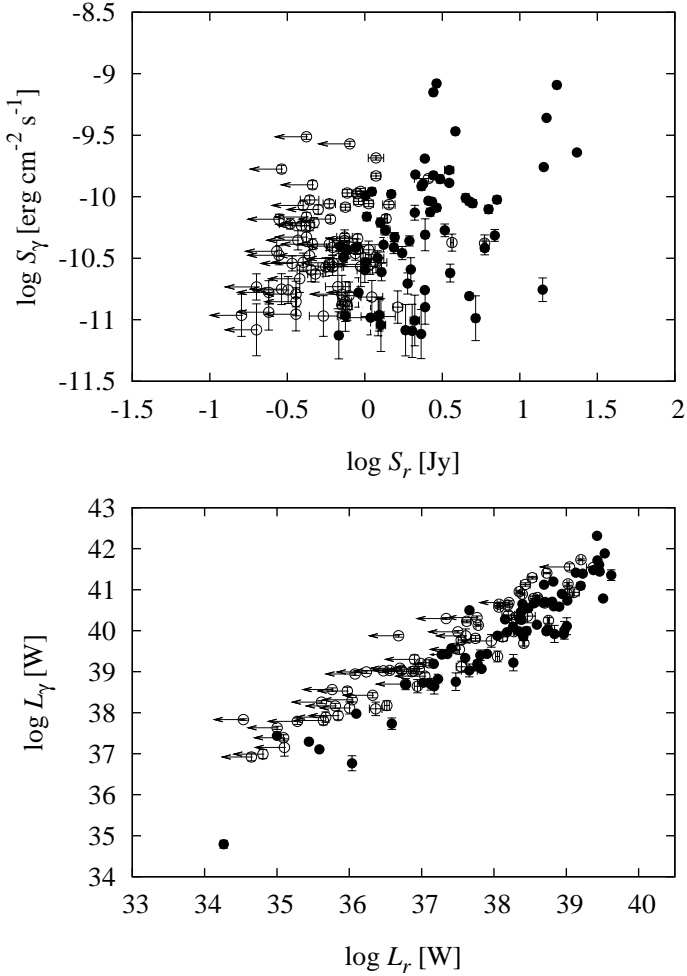}
\caption{The flux and luminosity correlations between 37 GHz and the \fermi band. Solid circles represent the complete northern sample, and 37 GHz upper limits are indicated by arrows.}
\label{fluxlum}
\end{figure}

In Fig.~\ref{fluxlum} we present the flux (top panel) and luminosity (bottom panel) correlations between the gamma and radio bands for the Mets\"ahovi sample. Many of the sources only have upper limit radio fluxes and luminosities. The outlier data point in the luminosity correlation is 3C 274 (M 87), having clearly lower luminosities in both bands but still fitting in the general trend. Significant correlation between the flux densities is found using the Kendall's tau test performed with the ASURV package. The flux correlation for the whole sample is strong with $\tau=0.195$ and $P<0.001$, where $\tau=0$ would indicate no correlation, and $P$ gives the probability of no correlation. Table~\ref{tableflux} lists the correlation coefficients for subsamples. Radio galaxies were omitted from the subsamples due to very small sample size. The flux correlation is significant for all subsamples except for BLOs and LPQs, and very strong for the northern sample. Although not listed in Table~\ref{tableflux}, the correlation for the whole sample is also significant ($\tau=0.230$ and $P<0.001$) when only detections are included and the upper limit datapoints discarded. 

It is evident in Fig.~\ref{fluxlum} that the errors in the gamma-ray flux densities increase towards the low-flux end. However, the errors in the 37 GHz flux densities are minimal, if at all visible. The net effect of errors for the shape of the correlation is small.

The appearance of the luminosity correlation is strikingly coherent. It is possible that it is in some measure created by the common dependence on redshift of the gamma and radio luminosities rather than their intrinsic correlation. Redshift bias and instrument flux limits are the two factors most likely to create spurious correlations \citep{elvis78, feigelson83,mucke97,bloom08}. In this study we have tried to alleviate their effect by including the upper limits in radio fluxes and luminosities and by using correlation tests taking the redshift bias into account. We calculated the significance of the luminosity correlation using the partial correlation method for censored data presented in \citet{akritas96}\footnote{Code by M. Akritas and J. Siebert is available at Penn State Center for Astrostatistics, http://astrostatistics.psu.edu/statcodes/index.html}. Even without the effect of redshift, the correlation is significant for the whole sample, complete northern sample, and all subsamples except for LPQs, which had a small sample size of 11 objects. The values of partial Kendall's $\tau$ and probabilites are listed in Table~\ref{tableflux}. 

\begin{table}
\caption{Partial Kendall's $\tau$ -coefficients and probabilities for no correlation between 37 GHz and gamma-ray fluxes and luminosities.}             
\label{tableflux}      
\centering                          
\begin{tabular}{l c c c c c c}        
\hline\hline 
 & \multicolumn{3}{c}{Flux correlation} &  \multicolumn{3}{c}{Luminosity correlation} \\               
Class & Number & $\tau$ & $P$ & Number & $\tau$ & $P$ \\    
\hline                        
  All & 141 & 0.195 & $<0.001$ & 140 & 0.288 & $<0.05$\\      
  BLO & 57 & 0.152 & 0.095 & 56 & 0.188 & $<0.05$\\
  HPQ & 34 & 0.414 & $<0.001$ & 34 & 0.426 & $<0.05$\\
  LPQ & 11 & 0.168 & 0.473 & 11 & 0.027 & $>0.05$\\ 
  QSO & 32 & 0.302 & 0.015 & 32 & 0.295 & $<0.05$\\ 
  CS\tablefootmark{a} & 67 & 0.310 & $<0.001$ & 67 & 0.433 & $<0.05$\\	
\hline                                   
\end{tabular}
\tablefoot{
\tablefoottext{a}{the complete northern sample}
}
\end{table}

It is noteworthy that the strength of both the flux and luminosity correlations is greatest for HPQs, then for QSOs and lowest for LPQs and BLOs. This effect was also noticed in \citet{leontavares11} using monthly-averaged flux densities in the same bands. We acknowledge the possibility that the correlation strengths and significances may be influenced by chance, especially when the sample size is small. To test the effect, we ran a set of Monte Carlo analyses for the correlations, randomly picking the same number of sources as in each of the subgroups from the whole sample. After a thousand iterations we determined the distributions of $\tau$ and $P$ for these random samples of $N$ sources. The simulation revealed that for LPQs the probability of not finding significant correlation in either flux or luminosity is high, irrespective of their true behaviour. Thus we cannot make any definite statements about their correlation one way or the other. However, the significance of both flux and luminosity correlation for HPQs is very high. The probability of finding such a correlation by chance is very low, meaning that the observed correlation in flux and luminosity is real. In flux density, the probability of finding a better correlation by chance is only 1\%. The situation for QSOs and BLOs is more complicated. In the case of flux density, the simulation results imply that we cannot say anything definite about them. For QSOs, the probability of finding a better correlation by chance is 21\%, which means that the correlation found in our study can also be spurious. Considering the luminosity correlation, it must be remembered that in the simulations the underlying distribution from which the random samples are picked is strongly correlated (the whole sample, $\tau=0.288$ and $P<0.05$). Testing correlations with intermediate strength, like those of QSOs and BLOs, is meaningless because when picking randomly a large fraction of the sources from a strongly correlated sample, the resulting sample is also very likely correlated to some degree.

\begin{figure}[]
\includegraphics[width=0.5\textwidth]{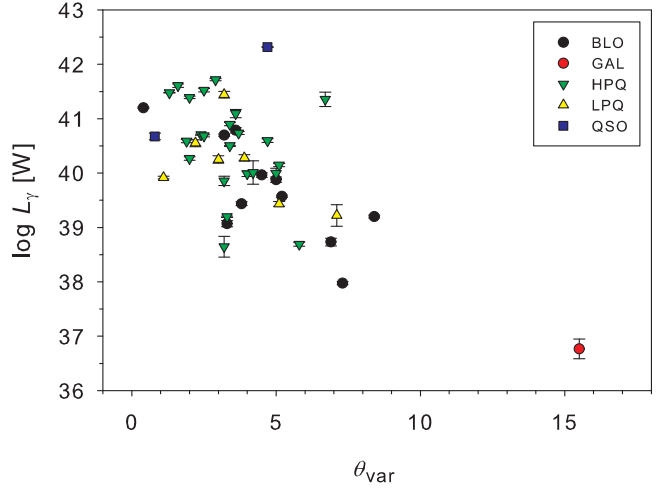}
\caption{Luminosity in the \fermi band plotted against the viewing angle. For clarity, sources 3C 84 (GAL, log\,$L_\gamma=37.1$\,W, $\theta_{var}=39.1$) and 3C 371.0 (BLO, log\,$L_\gamma=37.3$\,W, $\theta_{var}=57.3$) have been omitted from the figure.}
\label{lumtheta}
\end{figure}

As \citet{leontavares11} suspect, the difference in correlation strengths, if real, can be a result of the larger viewing angles of BLOs compared to QSOs and HPQs \citep{hovatta09}. Having a tighter flux and luminosity correlation for sources with extremely small viewing angles would agree well with the idea of gamma-rays originating in the jet. After all, when the jet is pointed towards us, the jet emission is all we see. We investigated this further by plotting the gamma-ray luminosity against the viewing angles from \citet{hovatta09}. The result is shown in Fig.~\ref{lumtheta}. We calculated the correlation with Spearman's rank correlation test with the Unistat v5.6.01 package. The correlation is significant for the whole sample ($\rho=-0.594$ and $P<0.001$). Only two of the data points do not belong in the complete northern sample, so the correlation is strong for the complete sample as well. For the subsamples we find a significant correlation for both BLOs ($\rho=-0.762$ and $P=0.002$) and HPQs ($\rho=-0.419$ and $P=0.023$). For LPQs the correlation is insignificant, possibly due to small sample size of 7 sources. 

\begin{figure}[]
\includegraphics[width=0.5\textwidth]{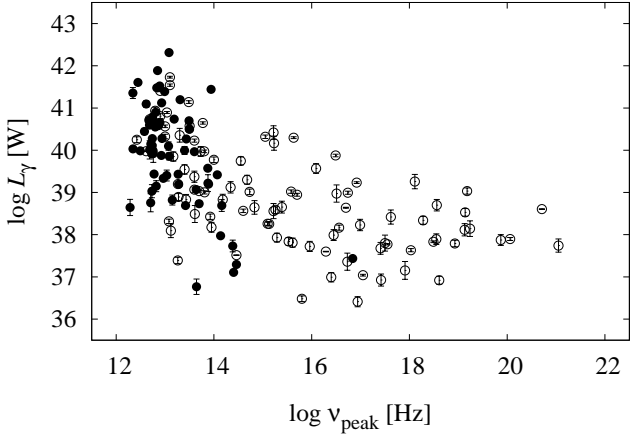}
\caption{The correlation between the gamma-ray luminosity, $L_\gamma$, and the peak frequency of the synchrotron component, $\nu_p$. Solid circles represent the complete northern sample.}
\label{gamlum_nu}
\end{figure}

Figure~\ref{gamlum_nu} shows the dependence between the gamma-ray luminosity, $L_\gamma$, and the peak frequency of the synchrotron component, log\,$\nu_p$. The peak frequencies were preferably taken from \citet{nieppola08} and then from \citet{nieppola06}. They are not from the same observational period as the \fermi data. However, because the log\,$\nu_p$ -values were calculated from averaged data, major shifts in log\,$\nu_p$ for a significant number of sources are unlikely. For 16 additional sources we took the peak frequency from \citet{abdo10_sed}, who use data from the 1FGL period. The Spearman correlation coefficients for the log\,$L_\gamma$ -- log\,$\nu_p$ correlation are $\rho=-0.706$ and $P<0.001$ for the whole sample and $\rho=-0.419$ and $P<0.001$ for the complete northern sample. Of the subsamples, the correlation is significant for BLOs and QSOs. When $\nu_p>14.5$, the sample consists of only BLOs, whose peak frequency range extends further into the high-energy region. As stated in \citet{nieppola06}, the values of $\nu_p$ may be exaggerated toward the high-$\nu_p$ end. However, this does not change the main shape of the correlation in Fig.~\ref{gamlum_nu}. The correlation would in fact only be stronger if the high-$\nu_p$ tail moved to lower frequencies.

\subsection{Gamma dominance}

\begin{figure}[]
\includegraphics[width=0.5\textwidth]{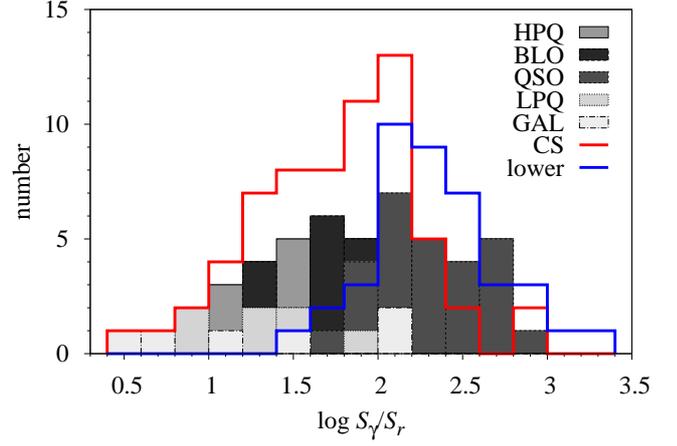}
\caption{The distribution of the gamma dominance log\,$S_\gamma/S_r$. Different AGN subsamples are depicted with different shades, and the coloured lines mark the complete sample and the sources with only lower limit estimate of the gamma dominance.}
\label{gamdom}
\end{figure}

In this paper we define gamma dominance as the ratio between the \fermilat $\gamma$-ray and 37 GHz radio fluxes, $S_\gamma/S_r$. We emphasize that what we call gamma dominance is different from Compton dominance, which equals the ratio of the peak luminosities of the IC and synchrotron components of the SED. Figure~\ref{gamdom} shows the distribution of the gamma dominance in our sample. Again, we drew the subsamples separately, and additionally plotted the distribution of the complete sample (red line). The gamma dominances of these samples were calculated using only the 37 GHz detections. For completeness, we have also plotted in the figure the distribution of the gamma dominances for the rest of the sample, which were calculated from upper limit average 37 GHz fluxes (resulting in lower limit gamma dominances, blue line in the figure). This lower limit sample contains BLOs for the most part (31/40), as well as five QSOs, two LPQs, and two GALs. Due to their low radio fluxes, the lower limit sources are located in the high end of the gamma dominance distribution. In contrast to the luminosity distributions (Fig.~\ref{radgamlum_hist}), all subsamples seem to have very similar distributions of gamma dominance. According to the Kruskal-Wallis test, QSOs are, however, signicantly different from others. The mean gamma dominance of QSOs is log\,$S_\gamma/S_r=2.3$ compared to log\,$S_\gamma/S_r=1.4-1.9$ of other subsamples.

\begin{figure}[]
\includegraphics[width=0.5\textwidth]{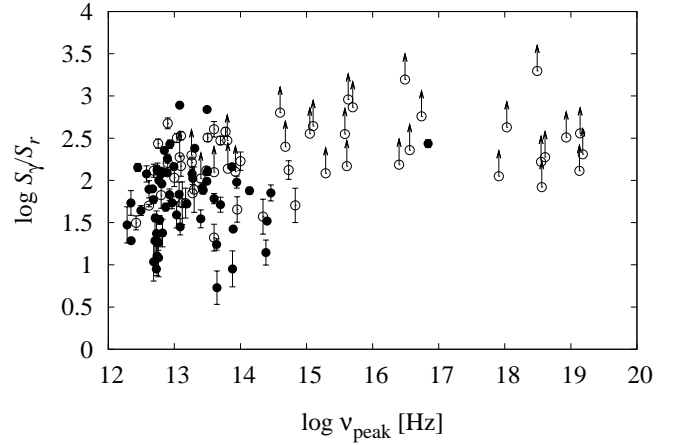}
\caption{The gamma dominance plotted against the synchrotron peak frequency. Solid circles represent the complete northern sample.}
\label{gamdom_nu}
\end{figure}

In Fig.~\ref{gamdom_nu} we have plotted the gamma dominance against the synchrotron peak frequency $\nu_p$. Similarly to Fig.~\ref{gamlum_nu}, objects beyond $\nu_p=14.5$ are exclusively radio-faint BLOs. In the whole sample there is a correlation ($\rho=0.537$ and $P<0.001$), but it disappears when only the complete sample is considered ($\rho=0.171$ and $P=0.173$). The positive correlation is also significant for BLOs, and just barely for the QSOs. 

\begin{figure}[]
\includegraphics[width=0.5\textwidth]{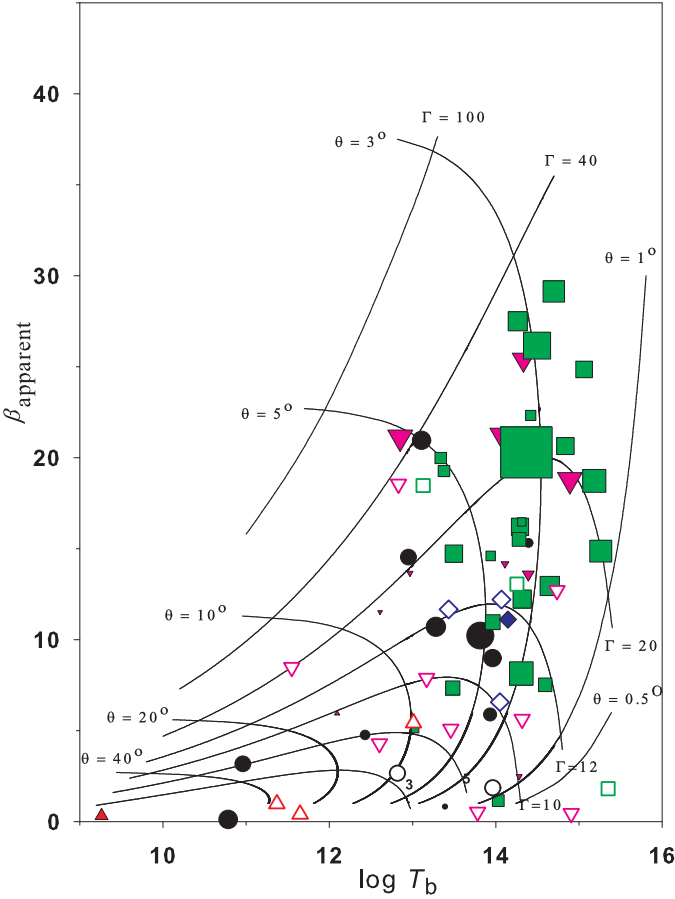}
\caption{The AGN sample of \citet{hovatta09} plotted as a function of apparent speed $\beta$, the brightness temperature $T_b$, the viewing angle $\theta$, and the Lorentz factor $\Gamma$. Open symbols are not detected by Fermi, filled symbols are detected, and the size of the symbol is proportional to the gamma dominance. Red triangles are galaxies, black circles BLOs, pink inverted triangles LPQs, green squares HPQs, and purple diamonds QSOs.}
\label{e_plot}
\end{figure}

In an effort to get hints of the main driver behind the gamma-ray emission, we incorporated several possible factors in the same plot. Fig.~\ref{e_plot} includes gamma dominances, the apparent velocity of the jet, the brightness temperature $T_b$, the viewing angle $\theta$, and the Lorentz factor of the jet. The figure in its original form was published in \citet{hovatta09}. We have added the gamma dominance to trace its dependence on the jet parameters. For completeness, we also included sources \textit{not} detected by \textit{Fermi}. 

The stronger gamma-ray dominance of QSOs is not evident in the figure because the sample size of QSOs is significantly reduced by the requirement of the jet parameters. This selection effect influences the whole sample in the figure, which must be remembered when interpreting it. The jet parameters are calculated from flares in the 22 or 37 GHz flux curves \citep{hovatta09}, so all results only apply for radio-bright, flaring sources and not categorically to all gamma-ray emitters. However, closer inspection reveals two things. The gamma-detected and -dominated sources have generally very small viewing angles and high brightness temperatures. Also at least the probability of gamma-ray detection grows with increasing $\Gamma$. However, there is no single property singling out the gamma-dominated sources. We conclude, like others before us \citep[e.g.,][]{lister09,savolainen10}, that jet parameters play an important role in the gamma-ray emission of radio-bright AGN.

\subsection{Variability}

It has been noted previously that AGN detected by \fermi are often flaring in radio or at least in an active state (i.e., variable) \citep{kovalev09,tornikoski10}. Recently, \citet{richards11} have carried out a detailed statistical study of differences in the variability amplitudes of gamma-ray detected and non-detected sources. They found that gamma-ray detected sources almost have a factor of two higher variability amplitude than sources not detected by \fermi.

\begin{figure}[]
\includegraphics[width=0.5\textwidth]{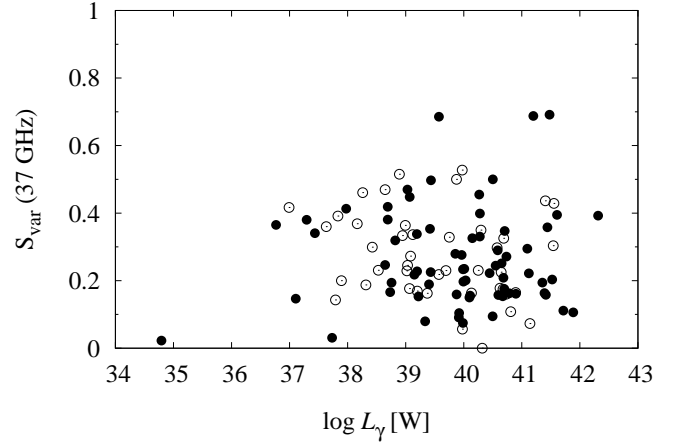}
\caption{The dependence between the variability at 37 GHz during the 1FGL period and the gamma-ray luminosity. Solid circles represent the complete northern sample.}
\label{var37_gam}
\end{figure}

However, among the gamma-ray detected sources, the brightest ones do not seem to be any more variable at 37 GHz during the 1FGL period than fainter ones. This is seen in Fig.~\ref{var37_gam} and confirmed by Spearman rank correlation test. The variability index at 37 GHz we used in Fig.~\ref{var37_gam} is \begin{equation}S_{var}=\frac{S_{max}-S_{min}}{S_{max}+S_{min}},\end{equation} where $S_{max}$ and $S_{min}$ are the maximum and minimum flux densities, respectively, during the variability period. However, the 1-year 1FGL period is a very short time when trying to illustrate AGN variability accurately. The variability index tends to get higher with increasing observing time, which is why we replotted the figure using the entire Mets\"ahovi database. The distribution of the datapoints remained very similar, but the population is moved toward higher $S_{var}$. Any significant correlation did not appear.

\section{Discussion}
\label{dis}

The undisputed correlation between the \fermilat $\gamma$-ray and 37 GHz radio luminosities presented in this paper (Fig.~\ref{fluxlum}) is a strong indication of the common origin of high radio frequency and gamma-ray radiation in AGN. The flux correlation, mostly at lower radio frequencies, has also been presented by other authors. \citet{giroletti10} find a dependence between the \fermilat and 8.4 Jy flux densities from the CRATES catalogue \citep{healey07}. \citet{mahony10} used the 20 GHz data (closer to ours in frequency) taken with the Australia Telescope Compact Array and, again, find a significant correlation. This 20 GHz correlation was studied in depth by \citet{ghirlanda10}, who use a numerical simulation to conclude that the flux correlation is actually real and not an effect of instrument flux limits. They also observe a tight luminosity correlation. In their data, the flux correlation for BLOs is significant, whereas in our 37 GHz data the correlation is rejected at the 5\% level. This is most likely due to the high number (almost half) of upper limits in our BLO radio fluxes, which may weaken the correlation. 

A significant luminosity correlation was also noted by \citet{bloom08} using \textit{CGRO}/EGRET data and applying Monte Carlo simulations. EGRET data were used earlier by \citet{mucke97}. They thoroughly studied the luminosity correlations through simulations and in practice between EGRET and Effelsberg 100m radio telescope data. In contrast to the later studies listed above, they found no correlation in any of the frequency bands (2.7, 4.8, 8, and 10 GHz). They considered simultaneous, maximum, and mean fluxes separately. One possible explanation for the correlation not appearing in their data is the use of relatively low radio frequencies. Below 15 GHz, the features in the flux curves are less distinct because the individual shock components are superposed. The emission typically is significantly delayed compared to the higher frequency bands, even 181 days at 4.8 GHz with respect to the flare peak at a higher radio frequency \citep{hovatta08}. Naturally, the correlation between low radio frequency flux and gamma-ray flux would be very difficult to detect, even if such a connection exists. The same effect is evident in \citet{richards10}, who compare \fermi data with simultaneous F-GAMMA project data. The flux correlation is tested with IRAM, Effelsberg, and OVRO data between 2.6 and 142 GHz. For frequencies above 10.5 GHz, the correlation is significant, but it cannot be confirmed for the lower frequencies.

In \citet{nieppola08} we observed that the inverse correlation between the synchrotron peak frequency and luminosity, a part of the blazar sequence scenario \citep{fossati97,ghisellini98,ghisellini08} and often observed for radio-bright blazars, can naturally be explained by the stronger Doppler boosting of low-$\nu_p$ sources. Once corrected for boosting, blazars show no correlation between the two quantities, and there is even an indication that high-$\nu_p$ blazars may have brighter synchrotron peaks in the rest frame. If we consider the correlation between radio and gamma-ray luminosities (Fig.~\ref{fluxlum}) and between gamma-ray luminosity and the viewing angle (Fig.~\ref{lumtheta}), it seems only natural that similar boosting also affects the gamma-ray emission. Stronger boosting of gamma-ray bright sources was also suspected by \citet{lahteenmaki03} and \fermi-detected AGN have been found more boosted than the non-detected by several authors \citep[e.g.,][]{kovalev09,savolainen10}. Therefore we consider it very possible that the correlation between $\nu_p$ and $L_\gamma$ would disappear if intrinsic values could be used. It is also well known that the limiting flux of the LAT telescope is a function of the photon spectral index \citep{abdo10_lac}. This means that it is more likely to detect faint sources with hard photon indices than with soft ones (see Fig.~9 of Abdo et al. 2010c). This probably explains a part of the different distributions of the subgroups in gamma luminosity, but does not account for the fact that sources of high \fermi luminosity are mostly quasars. 

Strong Doppler boosting in the \fermi-band would also offer an explanation for the stronger gamma dominance of high-$\nu_p$ sources (Fig.~\ref{gamdom_nu}). If the intrinsic synchrotron and Compton luminosities across the $\nu_p$-range were comparable, the \fermi luminosities for high-$\nu_p$ sources would be higher, on average. This is because the SED of low-$\nu_p$ sources is already falling in the \fermi band, while that of the high-$\nu_p$ sources is just peaking. This would result in stronger gamma-dominance in high-$\nu_p$ sources, as indicated. After all, being a ratio of fluxes, the gamma dominance is much less dependent on Doppler boosting, although spectral indices and the difference in Doppler factors in radio and gamma regions play a role.

\begin{figure}[]
\includegraphics[width=0.5\textwidth]{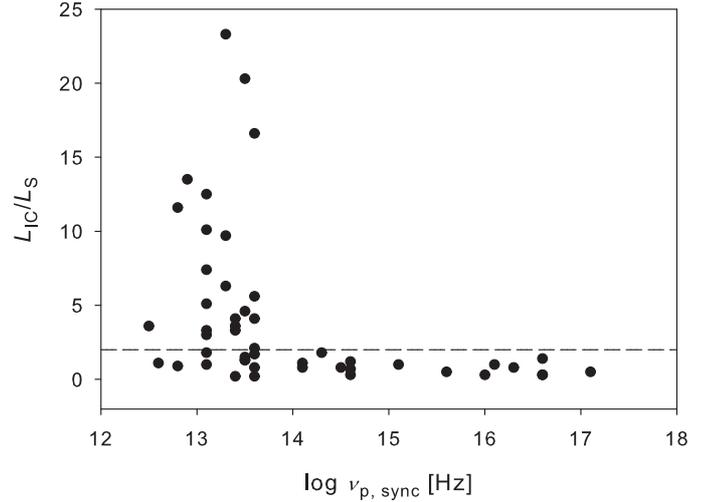}
\caption{The Compton dominance plotted against the synchrotron peak frequency, both data sets from \citet{abdo10_sed}. Sources above the dashed line at $L_{IC}/L_{s}=2$ are Compton dominated according to the classification of \citet{abdo10_sed}.} 
\label{abdo_cd}
\end{figure}

The role of Doppler boosting in the gamma region has not been studied with observational data owing to the difficulty of measuring the amount of boosting at high energies. The Compton dominance, $L_{IC}/L_{s}$, measured from robust SEDs for an extensive sample would also give a measure of the differences between the peak intensities. The best data set available for such a study is that of \citet{abdo10_sed}. There the Compton dominance is calculated for 48 sources. When plotted against log \,$\nu_p$, it is evident that the highest Compton dominances occur at low $\nu_p$ (Fig.~\ref{abdo_cd}). The low Compton dominances, in turn, are evenly spread in the $\nu_p$-range. \citet{abdo10_sed} define objects with $L_{IC}/L_{s}>2$ as Compton dominated. While all Compton dominated sources are low synchrotron-peaked blazars (LSP), they also form the biggest group among the non-Compton dominated sources. It seems that the correlation between $\nu_p$ and Compton dominance is far from simple. \citet{tramacere10} suggest that quasars with strong Compton domination have intrinsically different emission mechanism (EC) than sources (including LSP) with low Compton domination (SSC). Also \citet{lahteenmaki03} suspect that quasars and BLOs may have different mechanisms for gamma-ray emission. They, however, reach the conclusion that BLOs are weaker gamma emitters, and most likely at least a part of their emission is produced via the EC process, contrary to the scheme of \citet{tramacere10}. We hope the new data releases from \fermi will help in resolving these issues.

\section{Summary}

We have connected data from \fermilat 1FGL catalogue with the 37 GHz observations from Aalto University Mets\"ahovi Radio Observatory for 249 northern AGN through flux and luminosity correlations. We also calculated the gamma dominances and compared them, along with the gamma-ray luminosities, with the synchrotron peak frequency. From our results we draw the following summary:

\begin{enumerate}
\item The 37 GHz luminosities are significantly lower for BLOs than for quasar subsamples. A similar result for the \fermi luminosities has been found in earlier studies and corroborated by our data.

\item We find significant correlation between the \fermilat and 37 GHz flux densities for our whole sample ($\tau=0.195$ and $P<0.001$) and the complete northern sample ($\tau=0.310$ and $P<0.001$), as well as HPQ and QSOs. For BLOs and LPQs the correlation is not significant. 

\item We find a significant correlation for \fermilat and 37 GHz luminosities for the whole sample ($\tau=0.288$ and $P<0.05$) and all subsamples except for LPQs.

\item There is a strong negative correlation between the viewing angle and the average \fermi luminosity, for the whole sample, as well as for BLOs and HPQs alone.

\item The gamma-ray luminosities of high-synchrotron peaked blazars are lower than those of low-synchrotron peaked blazars, but high-peaked blazars seem to be more gamma-ray dominated when using the \fermilat and 37 GHz bands. This could be explained with the significantly stronger Doppler boosting of low-synchrotron peaked blazars in the gamma region.

\item The long-term 37 GHz variability does not depend on the gamma-ray luminosity; i.e., high \fermi luminosity objects are not more variable at 37 GHz.

\item As found in previous studies, we also conclude that BLOs are different gamma-ray emitters than quasars. This may be the result of stronger Doppler boosting for quasars or of inherently different emission mechanisms.

\item Our results support the view that gamma-rays are produced co-spatially with the 37 GHz radiation, i.e., in the jet, mainly through the SSC process. Co-spatial origin of radio and gamma-ray emission would also serve as a natural explanation of the possibly similar relativistic boosting properties in the two bands.
\end{enumerate}

\begin{acknowledgements}
The Mets\"ahovi team acknowledges the support from the Academy of Finland
for our observing projects (numbers 212656, 210338, 121148, and others).
\end{acknowledgements}


\bibliographystyle{aa}
\bibliography{/home/eni/vip/enbib}

\Online

\longtab{1}{
\begin{longtable}{l l c c c c c c c}   
\caption{The sample sources used in this study and their classifications. }\label{sample} \\  
\noalign{\smallskip}               
\hline\hline          
\noalign{\smallskip}
Name & 1FGL name & Class & Complete sample & R.A.(J2000) & Dec(J2000) & $z$ & $S_r$ & log\,$S_\gamma$\\
\noalign{\smallskip}
 & & & & & & & $[\textrm{Jy}]$ & $[\textrm{erg cm}^{-2} \textrm{s}^{-1}]$\\ 
\noalign{\smallskip}
\hline 
\noalign{\smallskip}    
\endfirsthead        
\caption{continued.}\\
\noalign{\smallskip}
\hline\hline
\noalign{\smallskip}
Name & 1FGL name & Class & Complete sample & R.A.(J2000) & Dec(J2000) & $z$ & $S_r$ & log\,$S_\gamma$\\
\noalign{\smallskip}
 & & & & & & & $[\textrm{Jy}]$ & $[\textrm{erg cm}^{-2} \textrm{s}^{-1}]$\\ 
\noalign{\smallskip}
\hline
\noalign{\smallskip}
\endhead   
\hline
\endfoot   
CGRaBS J0017--0512   &   J0017.4--0510    &   QSO   & ... & 4.37 & --5.18 & ... & 0.42 & ... \\
RX J0018.4+2947   &   J0018.6+2945    &   BLO   & ... & 4.65 & 29.76 & 0.10 & ... & --10.80 \\
PKS 0017+200   &   J0019.3+2017    &   BLO   & ... & 4.83 & 20.29 & ... & ... & ... \\
PKS 0019+058   &   J0022.5+0607    &   BLO   & ... & 5.63 & 6.13 & ... & 0.28 & ... \\
RX J0035.2+1515   &   J0035.1+1516    &   BLO   & ... & 8.79 & 15.27 & 1.09 & ... & --10.86 \\
1ES 0033+595   &   J0035.9+5951    &   BLO   & ... & 8.99 & 59.85 & 0.09 & $<0.28$ & --10.48 \\
0039+230   &   J0041.9+2318    & ... & ... & 10.48 & 23.31 & 1.43 & ... & --10.75 \\
RX J0045.3+2127   &   J0045.3+2127    &   BLO   & ... & 11.34 & 21.46 & ... & ... & ... \\
PKS 0047+023   &   J0050.2+0235    &   BLO   & ... & 12.55 & 2.59 & ... & ... & ... \\
0048--097   &   J0050.6--0928    &   BLO   &   C   & 12.66 & --9.48 & 0.30 & 1.36 & --10.27 \\
PKS 0048--071   &   J0051.1--0649    &   QSO   & ... & 12.78 & --6.82 & 1.97 & 1.18 & --10.53 \\
J0100+0745   &   J0100.2+0747    & ... & ... & 15.05 & 7.80 & ... & ... & ... \\
0059+581   &   J0102.8+5827    &   HPQ   &   C   & 15.71 & 58.47 & 0.64 & 3.28 & --10.27 \\
0106+013   &   J0108.6+0135    &   HPQ   &   C   & 17.17 & 1.59 & 2.10 & 2.12 & --9.82 \\
RGB J0109+182   &   J0109.0+1816    &   BLO   & ... & 17.26 & 18.28 & 0.15 & ... & --11.16 \\
S2 0109+22   &   J0112.0+2247    &   BLO   & ... & 18.02 & 22.79 & 0.27 & $<0.9$ & --10.34 \\
4C 31.03   &   J0112.9+3207    &   QSO   & ... & 18.23 & 32.12 & 0.60 & 1.06 & --10.06 \\
RX J0115.7+2519   &   J0115.5+2519    &   BLO   & ... & 18.89 & 25.33 & 0.37 & ... & --10.85 \\
B3 0133+388   &   J0136.5+3905    &   BLO   & ... & 24.13 & 39.09 & ... & $<0.56$ & ... \\
0133+476   &   J0137.0+4751    &   HPQ   &   C   & 24.25 & 47.85 & 0.86 & 3.50 & --9.89 \\
PKS 0139--09   &   J0141.7--0929    &   BLO   & ... & 25.43 & --9.49 & 0.73 & $<0.36$ & --10.85 \\
J0144+2705   &   J0144.6+2703    & ... & ... & 26.17 & 27.06 & ... & ... & ... \\
RX J0159.5+1047  &   J0159.5+1047    &   BLO   & ... & 29.89 & 10.80 & ... & ... & ... \\
RX J0202.4+0849   &   J0202.1+0849    &   BLO   & ... & 30.53 & 8.83 & ... & ... & ... \\
S5 0159+72   &   J0203.5+7234    &   BLO   & ... & 30.89 & 72.57 & ... & $<0.48$ & ... \\
0202+149   &   J0204.5+1516    &   HPQ   &   C   & 31.13 & 15.28 & 0.41 & 0.68 & --11.13 \\
MS 0205.7+3509   &   J0208.6+3522    &   BLO   & ... & 32.17 & 35.38 & 0.32 & ... & --10.97 \\
J0211+1051   &   J0211.2+1049    &   QSO   & ... & 32.81 & 10.83 & ... & 0.71 & ... \\
Zel 0214+083   &   J0217.2+0834    &   BLO   & ... & 34.32 & 8.58 & 1.40 & ... & --10.93 \\
0212+735   &   J0217.8+7353    &   HPQ   &   C   & 34.45 & 73.88 & 2.37 & 2.45 & --10.31 \\
0215+015   &   J0217.9+0144    &   HPQ   & ... & 34.48 & 1.75 & 1.72 & 1.37 & --10.18 \\
0218+357   &   J0221.0+3555    &   BLO   & ... & 35.27 & 35.93 & 0.94 & 0.91 & --10.04 \\
3C 66A   &   J0222.6+4302    &   BLO   & ... & 35.67 & 43.04 & 0.44 & $<0.8$ & --9.57 \\
0234+285   &   J0237.9+2848    &   HPQ   &   C   & 39.49 & 28.80 & 1.21 & 2.71 & --10.04 \\
0235+164   &   J0238.6+1637    &   BLO   &   C   & 39.67 & 16.62 & 0.94 & 3.84 & --9.47 \\
S5 0238+71   &   J0243.5+7116    &   BLO   & ... & 40.89 & 71.28 & ... & $<0.4$ & ... \\
RX J0250.6+1712   &   J0250.4+1715    &   BLO   & ... & 42.61 & 17.26 & 1.10 & $<0.16$ & --10.96 \\
4C 47.08   &   J0303.1+4711    &   BLO   & ... & 45.79 & 47.19 & 0.48 & 1.11 & --10.82 \\
RX J0316.1+0904   &   J0316.1+0904    &   BLO   & ... & 49.05 & 9.08 & ... & ... & ... \\
MS 0317.0+1834   &   J0319.7+1847    &   BLO   & ... & 49.93 & 18.80 & 0.19 & ... & --10.79 \\
3C 84   &   J0319.7+4130    &   GAL   &   C   & 49.94 & 41.51 & 0.02 & 14.28 & --9.76 \\
RGB J0321+2336   &   J0322.1+2336    &   BLO   & ... & 50.54 & 23.61 & ... & ... & ... \\
2E 0323+0214   &   J0326.2+0222    &   BLO   & ... & 51.57 & 2.38 & 0.15 & ... & --10.89 \\
CTA 026   &   J0339.2--0143    &   HPQ   &   C   & 54.80 & --1.72 & 0.85 & 1.89 & --10.70 \\
2E 0414+0057   &   J0416.8+0107    &   BLO   & ... & 64.21 & 1.12 & 0.29 & ... & --10.81 \\
0415+379   &   J0419.0+3811    &   GAL   &   C   & 64.76 & 38.19 & 0.05 & 5.19 & --10.99 \\
PKS 0420+022   &   J0422.1+0211    &   BLO   & ... & 65.53 & 2.20 & 2.28 & 0.67 & --10.73 \\
0420--014   &   J0423.2--0118    &   HPQ   &   C   & 65.80 & --1.31 & 0.92 & 4.96 & --10.05 \\
PKS 0422+0036   &   J0424.8+0036    &   BLO   & ... & 66.21 & 0.61 & 0.31 & 0.75 & --10.85 \\
2EG J0432+2910   &   J0433.5+2905    &   BLO   & ... & 68.39 & 29.09 & ... & $<0.44$ & ... \\
NRAO 190   &   J0442.7--0019    &   QSO   & ... & 70.69 & --0.32 & 0.84 & 0.89 & --9.98 \\
PKS 0446+112   &   J0448.6+1118    &   GAL   &   C   & 72.17 & 11.31 & 1.21 & 0.73 & --10.49 \\
PKS 0458--020   &   J0501.0--0200    &   HPQ   &   C   & 75.27 & --2.00 & 2.29 & 1.01 & --10.53 \\
RX J0505.5+0416   &   J0505.2+0420    &   BLO   & ... & 76.31 & 4.34 & 0.03 & ... & --10.81 \\
1ES 0502+675   &   J0507.9+6738    &   BLO   & ... & 76.99 & 67.64 & 0.31 & $<0.44$ & --10.48 \\
0506+101   &   J0509.2+1015    & ... & ... & 77.31 & 10.26 & ... & 0.64 & ... \\
MG 0509+0541   &   J0509.3+0540    &   BLO   & ... & 77.34 & 5.67 & ... & ... & ... \\
0507+179   &   J0510.0+1800    &   HPQ   &   C   & 77.50 & 18.02 & 0.42 & 1.27 & --11.04 \\
0528+134   &   J0531.0+1331    &   HPQ   &   C   & 82.75 & 13.52 & 2.07 & 2.32 & --9.91 \\
0539--057   &   J0540.9--0547    &   LPQ   & ... & 85.23 & --5.79 & 0.84 & $<0.71$ & --10.79 \\
PKS 0605--085   &   J0608.2--0837    &   HPQ   &   C   & 92.05 & --8.62 & 0.87 & 1.98 & --10.59 \\
0621+446   &   J0625.4+4440    &   BLO   & ... & 96.36 & 44.67 & ... & $<0.36$ & ... \\
1ES 0647+250   &   J0650.7+2503    &   BLO   & ... & 102.68 & 25.06 & 0.20 & ... & --10.74 \\
B3 0650+453   &   J0654.3+4514    &   QSO   & ... & 103.58 & 45.25 & 0.93 & ... & --10.09 \\
J0654+5042   &   J0654.4+5042    & ... & ... & 103.62 & 50.71 & ... & $<0.39$ & ... \\
EXO 0706.1+591   &   J0710.6+5911    &   BLO   & ... & 107.66 & 59.19 & 0.13 & ... & --10.87 \\
J0712+5033   &   J0712.7+5033    &   BLO   & ... & 108.18 & 50.56 & ... & $<0.56$ & ... \\
J0713+1935   &   J0714.0+1935    & ... & ... & 108.51 & 19.59 & ... & 0.70 & ... \\
0716+332   &   J0719.3+3306    &   QSO   & ... & 109.85 & 33.11 & 0.78 & 0.39 & --10.24 \\
0718+042   &   J0721.4+0401    & ... & ... & 110.35 & 4.03 & ... & 0.73 & ... \\
0716+714   &   J0721.9+7120    &   BLO   &   C   & 110.48 & 71.34 & 0.30 & 2.38 & --9.89 \\
RX J0723.2+5841   &   J0722.3+5837    &   BLO   & ... & 110.59 & 58.63 & ... & ... & ... \\
PKS 0723--008   &   J0725.9--0053    &   GAL   &   C   & 111.50 & --0.90 & 0.13 & 2.45 & --10.90 \\
FBQS J0730+3307   &   J0730.0+3305    &   BLO   & ... & 112.52 & 33.09 & 0.11 & ... & --10.90 \\
PKS 0735+17   &   J0738.2+1741    &   BLO   &   C   & 114.56 & 17.70 & 0.42 & $<0.69$ & --10.40 \\
0736+017   &   J0739.1+0138    &   HPQ   &   C   & 114.79 & 1.64 & 0.19 & 1.56 & --10.33 \\
0738+5451   &   J0742.2+5443    &   QSO   & ... & 115.56 & 54.73 & 0.72 & 0.62 & --10.57 \\
MS 0737.9+7441   &   J0745.2+7438    &   BLO   & ... & 116.32 & 74.64 & 0.32 & ... & --11.02 \\
J0746.3+2548   &   J0746.6+2548    &   QSO   & ... & 116.66 & 25.80 & 2.98 & $<0.37$ & --10.35 \\
0748+126   &   J0750.6+1235    &   LPQ   &   C   & 117.67 & 12.59 & 0.89 & 3.55 & --10.62 \\
S4 0749+54   &   J0752.8+5353    &   BLO   & ... & 118.20 & 53.89 & 0.20 & 0.54 & --10.97 \\
0754+100   &   J0757.2+0956    &   BLO   &   C   & 119.31 & 9.94 & 0.27 & 1.28 & --10.61 \\
RX J0805.4+7534   &   J0804.7+7534    &   BLO   & ... & 121.19 & 75.57 & 0.12 & ... & --10.86 \\
0805--077   &   J0808.2--0750    &   QSO   &   C   & 122.05 & --7.84 & 1.84 & 1.48 & --9.98 \\
1ES 0806+524   &   J0809.5+5219    &   BLO   & ... & 122.39 & 52.32 & 0.14 & $<0.34$ & --10.54 \\
PKS 0808+019   &   J0811.2+0148    &   BLO   & ... & 122.81 & 1.82 & 0.93 & 0.79 & --10.80 \\
RX J0816.3+5739   &   J0816.7+5739    &   BLO   & ... & 124.19 & 57.66 & ... & ... & ... \\
0814+425   &   J0818.2+4222    &   BLO   & ... & 124.55 & 42.38 & 0.25 & 1.43 & --10.06 \\
0820+560   &   J0825.0+5555    &   HPQ   & ... & 126.27 & 55.93 & 1.42 & 0.74 & --10.33 \\
0823+033   &   J0825.9+0309    &   HPQ   &   C   & 126.49 & 3.15 & 0.51 & 1.22 & --10.97 \\
OJ 248   &   J0830.5+2407    &   LPQ   &   C   & 127.63 & 24.12 & 0.94 & $<1.32$ & --10.39 \\
0829+046   &   J0831.6+0429    &   BLO   & ... & 127.91 & 4.49 & 0.18 & $<0.62$ & --10.54 \\
0836+710   &   J0842.2+7054    &   LPQ   &   C   & 130.57 & 70.90 & 2.17 & 2.10 & --10.13 \\
RX J0847.2+1133   &   J0847.2+1134    &   BLO   & ... & 131.82 & 11.57 & 0.20 & $<0.24$ & --10.94 \\
J0850--1213   &   J0850.0--1213    &   QSO   & ... & 132.52 & --12.22 & 0.57 & 0.59 & --10.55 \\
OJ 287   &   J0854.8+2006    &   BLO   &   C   & 133.71 & 20.11 & 0.31 & 5.94 & --10.42 \\
Ton 1015   &   J0910.7+3332    &   BLO   & ... & 137.68 & 33.54 & 0.35 & ... & --10.97 \\
B2 0912+29   &   J0915.7+2931    &   BLO   & ... & 138.94 & 29.53 & ... & ... & ... \\
0917+449   &   J0920.9+4441    &   QSO   &   C   & 140.24 & 44.69 & 2.18 & 2.44 & --9.69 \\
J0948+0022   &   J0949.0+0021    &   GAL   & ... & 147.25 & 0.36 & 0.58 & $<0.6$ & --10.18 \\
0953+254   &   J0956.9+2513    &   LPQ   &   C   & 149.24 & 25.22 & 0.71 & 0.75 & --10.97 \\
S4 0954+556   &   J0957.7+5523    &   HPQ   & ... & 149.43 & 55.39 & 0.90 & 0.77 & --9.97 \\
S4 0954+65   &   J1000.1+6539    &   BLO   &   C   & 150.03 & 65.65 & 0.37 & $<1.09$ & --10.98 \\
EXO 1004.0+350   &   J1007.0+3454    &   BLO   & ... & 151.77 & 34.90 & 0.61 & ... & --10.97 \\
NRAO 350   &   J1012.2+0634    &   BLO   & ... & 153.06 & 6.57 & 0.73 & ... & --10.82 \\
GB 1011+496   &   J1015.1+4927    &   BLO   & ... & 153.79 & 49.45 & 0.20 & $<0.4$ & --10.07 \\
1013+054   &   J1016.1+0514    &   QSO   & ... & 154.03 & 5.24 & 1.71 & 0.44 & --10.03 \\
RX J1022.7--0112   &   J1022.8--0115    &   BLO   & ... & 155.72 & --1.25 & ... & ... & ... \\
1ES 1028+511   &   J1031.0+5051    &   BLO   & ... & 157.76 & 50.86 & 0.36 & $<0.36$ & --10.96 \\
B3 1029+378   &   J1032.7+3737    &   BLO   & ... & 158.18 & 37.63 & ... & ... & ... \\
J 1033+6051   &   J1033.8+6048    &   QSO   & ... & 158.47 & 60.80 & ... & ... & ... \\
RX J1037.7+5711   &   J1037.7+5711    &   BLO   & ... & 159.44 & 57.20 & ... & ... & ... \\
TXS 1040+244   &   J1043.1+2404    &   BLO   &   C   & 160.79 & 24.08 & 0.56 & 1.24 & --10.96 \\
MS 1050.7+4946   &   J1053.6+4927    &   BLO   & ... & 163.42 & 49.45 & 0.14 & $<0.36$ & --10.79 \\
J1054+2210   &   J1054.5+2212    & ... & ... & 163.63 & 22.21 & ... & $<0.16$ & ... \\
1055+018   &   J1058.4+0134    &   HPQ   &   C   & 164.62 & 1.58 & 0.89 & 4.48 & --10.01 \\
1055+567   &   J1058.6+5628    &   QSO   & ... & 164.67 & 56.48 & 0.14 & $<0.28$ & --10.18 \\
Mrk 421   &   J1104.4+3812    &   BLO   & ... & 166.12 & 38.21 & 0.03 & $<0.42$ & --9.51 \\
RX J1110.6+7133   &   J1109.9+7134    &   BLO   & ... & 167.50 & 71.58 & ... & ... & ... \\
RX J1117.0+2014  &   J1117.1+2013    &   BLO   & ... & 169.30 & 20.22 & 0.10 & ... & --10.64 \\
EXO 1118.0+422   &   J1121.0+4209    &   BLO   & ... & 170.25 & 42.16 & 0.12 & ... & --10.93 \\
1118--056   &   J1121.5--0554    &   QSO   & ... & 170.38 & --5.91 & 1.30 & $<0.42$ & --10.33 \\
RX J1136.5+6737   &   J1136.2+6739    &   BLO   & ... & 174.06 & 67.66 & 0.14 & ... & --10.92 \\
Mrk 180   &   J1136.6+7009    &   BLO   & ... & 174.16 & 70.16 & 0.05 & $<0.24$ & --10.78 \\
RX J1136.8+2551    &   J1136.9+2551    &   BLO   & ... & 174.24 & 25.86 & 0.20 & ... & --10.81 \\
B2 1147+24   &   J1150.2+2419    &   BLO   & ... & 177.57 & 24.32 & 0.20 & 0.77 & --10.89 \\
RX J1151.4+5859   &   J1151.6+5857    &   BLO   & ... & 177.91 & 58.96 & ... & ... & ... \\
4C 29.45   &   J1159.4+2914    &   HPQ   &   C   & 179.85 & 29.24 & 0.73 & 2.64 & --10.13 \\
B3 1206+416   &   J1209.4+4119    &   BLO   & ... & 182.37 & 41.32 & ... & $<0.36$ & ... \\
B2 1215+30   &   J1217.7+3007    &   BLO   & ... & 184.45 & 30.12 & 0.24 & $<0.47$ & --10.21 \\
GB2 1217+348   &   J1220.2+3432    &   BLO   & ... & 185.06 & 34.53 & 0.13 & ... & --11.14 \\
PG 1218+304   &   J1221.3+3008    &   BLO   & ... & 185.34 & 30.14 & 0.18 & $<0.27$ & --10.44 \\
ON 231   &   J1221.5+2814    &   BLO   & ... & 185.39 & 28.25 & 0.10 & $<0.42$ & --10.17 \\
1219+044   &   J1222.5+0415    &   QSO   &   C   & 185.64 & 4.27 & 0.97 & 1.00 & --10.60 \\
PKS 1222+216   &   J1224.7+2121    &   LPQ   &   C   & 186.20 & 21.36 & 0.44 & 0.87 & --10.41 \\
S5 1221+80   &   J1224.8+8044    &   BLO   & ... & 186.22 & 80.74 & ... & ... & ... \\
3C 273   &   J1229.1+0203    &   HPQ   &   C   & 187.28 & 2.05 & 0.16 & 23.35 & --9.64 \\
3C 274   &   J1230.8+1223    &   GAL   &   C   & 187.72 & 12.39 & 0.00 & 14.03 & --10.76 \\
B2 1229+29   &   J1231.6+2850    &   BLO   & ... & 187.90 & 28.83 & 1.00 & ... & --10.60 \\
1237+0459   &   J1239.5+0443    &   QSO   & ... & 189.89 & 4.72 & 1.76 & 0.33 & --10.22 \\
Ton 116   &   J1243.1+3627    &   BLO   & ... & 190.79 & 36.46 & 1.07 & ... & --10.63 \\
PG 1246+586   &   J1248.2+5820    &   BLO   & ... & 192.06 & 58.34 & 0.85 & $<0.44$ & --10.23 \\
S4 1250+53   &   J1253.0+5301    &   BLO   & ... & 193.26 & 53.02 & ... & $<0.4$ & ... \\
3C 279   &   J1256.2--0547    &   HPQ   &   C   & 194.05 & --5.79 & 0.54 & 14.87 & --9.36 \\
MC2 1307+12.1   &   J1309.2+1156    &   BLO   & ... & 197.31 & 11.94 & ... & ... & ... \\
1308+326   &   J1310.6+3222    &   BLO   &   C   & 197.65 & 32.37 & 1.00 & 2.57 & --10.03 \\
TXS 1312+240   &   J1314.7+2346    &   BLO   & ... & 198.68 & 23.77 & ... & ... & ... \\
1324+224   &   J1326.6+2213    &   QSO   &   C   & 201.65 & 22.22 & 1.40 & 0.77 & --10.43 \\
PKS 1329--049   &   J1331.9--0506    &   QSO   & ... & 202.99 & --5.11 & 2.15 & 1.18 & --9.83 \\
J1333+5057   &   J1333.2+5056    & ... & ... & 203.30 & 50.95 & ... & 0.24 & ... \\
1334--127   &   J1337.7--1255    &   LPQ   & ... & 204.43 & --12.93 & 0.54 & 5.93 & --10.38 \\
RX J1340.4+4410   &   J1340.6+4406    &   BLO   & ... & 205.17 & 44.12 & 0.54 & ... & --11.10 \\
B2 1338+40   &   J1341.3+3951    &   BLO   & ... & 205.33 & 39.85 & 0.16 & ... & --10.97 \\
PKS 1352--104   &   J1354.9--1041    &   QSO   & ... & 208.73 & --10.69 & 0.33 & $<0.68$ & --10.50 \\
1357+769   &   J1358.1+7646    &   QSO   & ... & 209.53 & 76.77 & ... & $<0.4$ & ... \\
PKS 1406--076   &   J1408.9--0751    &   QSO   & ... & 212.23 & --7.85 & 1.49 & 0.76 & --10.44 \\
2E 1415+2557   &   J1417.8+2541    &   BLO   & ... & 214.46 & 25.68 & 0.24 & ... & --11.09 \\
RX J1422.6+5801   &   J1422.2+5757    &   BLO   & ... & 215.56 & 57.96 & 0.64 & ... & --10.97 \\
RX J1426.1+3404    &   J1426.0+3403    &   BLO   & ... & 216.52 & 34.06 & ... & ... & ... \\
PKS 1424+240   &   J1426.9+2347    &   BLO   & ... & 216.75 & 23.80 & 0.16 & $<0.46$ & --9.90 \\
H 1426+428   &   J1428.7+4239    &   BLO   & ... & 217.18 & 42.66 & 0.13 & $<0.29$ & --10.75 \\
RX J1436.9+5639   &   J1437.0+5640    &   BLO   & ... & 219.26 & 56.67 & 0.15 & ... & --10.99 \\
PG 1437+398   &   J1439.2+3930    &   BLO   & ... & 219.81 & 39.50 & 0.35 & ... & --10.98 \\
1ES 1440+122   &   J1442.8+1158    &   BLO   & ... & 220.71 & 11.97 & 0.16 & ... & --10.87 \\
RX J1448.0+3608   &   J1447.9+3608    &   BLO   & ... & 221.99 & 36.15 & ... & ... & ... \\
SBS 1452+516   &   J1454.6+5125    &   BLO   & ... & 223.66 & 51.42 & 1.08 & ... & --10.64 \\
MS 1458.8+2249   &   J1501.1+2237    &   BLO   & ... & 225.28 & 22.62 & 0.24 & ... & --10.63 \\
PKS 1502+106   &   J1504.4+1029    &   QSO   &   C   & 226.10 & 10.49 & 1.84 & 2.90 & --9.08 \\
1502+036   &   J1505.0+0328    &   QSO   & ... & 226.27 & 3.47 & 0.41 & 0.56 & --10.57 \\
SBS 1508+561   &   J1509.4+5602    &   BLO   & ... & 227.35 & 56.04 & 1.68 & ... & --10.88 \\
PKS 1508--055   &   J1511.1--0545    &   QSO   & ... & 227.79 & --5.76 & 1.19 & 0.94 & --10.57 \\
PKS 1510--089   &   J1512.8--0906    &   HPQ   &   C   & 228.21 & --9.10 & 0.36 & 2.77 & --9.15 \\
PKS 1514+197   &   J1516.9+1928    &   BLO   & ... & 229.25 & 19.48 & 0.65 & 1.63 & --10.90 \\
1H 1515+660   &   J1517.8+6530    &   BLO   & ... & 229.46 & 65.51 & 0.70 & ... & --11.09 \\
J1522+3144   &   J1522.1+3143    &   QSO   & ... & 230.55 & 31.73 & ... & $<0.4$ & ... \\
RX J1532.0+3016   &   J1531.8+3018    &   BLO   & ... & 232.96 & 30.31 & 0.06 & ... & --10.95 \\
1541+8204   &   J1536.6+8200    & ... & ... & 234.15 & 82.01 & ... & ... & ... \\
RX J1542.9+6129   &   J1542.9+6129    &   BLO   & ... & 235.74 & 61.49 & ... & ... & ... \\
1546+027   &   J1549.3+0235    &   HPQ   &   C   & 237.34 & 2.60 & 0.41 & 1.74 & --10.46 \\
1548+056   &   J1550.7+0527    &   HPQ   &   C   & 237.68 & 5.46 & 1.42 & 2.10 & --11.01 \\
PKS 1551+130   &   J1553.4+1255    &   QSO   & ... & 238.36 & 12.93 & 1.29 & 0.59 & --10.06 \\
PG 1553+11   &   J1555.7+1111    &   BLO   & ... & 238.94 & 11.19 & 0.36 & $<0.29$ & --9.77 \\
MYC 1557+566   &   J1558.9+5627    &   BLO   & ... & 239.74 & 56.46 & 0.30 & ... & --10.73 \\
PKS 1604+159   &   J1607.1+1552    &   BLO   & ... & 241.79 & 15.87 & 0.36 & 0.48 & --10.63 \\
1606+106   &   J1609.0+1031    &   LPQ   &   C   & 242.25 & 10.53 & 1.23 & 0.89 & --10.41 \\
DA 406   &   J1613.5+3411    &   HPQ   &   C   & 243.39 & 34.19 & 1.40 & 2.02 & --11.09 \\
4C 38.41   &   J1635.0+3808    &   HPQ   &   C   & 248.77 & 38.14 & 1.81 & 3.06 & --9.86 \\
3C 345   &   J1642.5+3947    &   HPQ   &   C   & 250.64 & 39.79 & 0.59 & 7.12 & --10.02 \\
Mrk 501   &   J1653.9+3945    &   BLO   &   C   & 253.49 & 39.75 & 0.03 & 1.01 & --9.99 \\
J1700+6830   &   J1700.1+6830    &   QSO   & ... & 255.05 & 68.51 & 0.30 & $<0.46$ & --10.39 \\
PKS 1717+177   &   J1719.2+1745    &   BLO   & ... & 259.81 & 17.76 & 0.14 & $<0.59$ & --10.39 \\
B 21722+40   &   J1724.0+4002    &   BLO   & ... & 261.01 & 40.04 & 1.05 & 0.87 & --10.46 \\
H 1722+119   &   J1725.0+1151    &   BLO   & ... & 261.27 & 11.86 & 0.02 & ... & --10.38 \\
ZW I 187   &   J1727.9+5010    &   BLO   & ... & 262.00 & 50.18 & 0.06 & ... & --10.93 \\
PKS 1725+044   &   J1728.2+0431    &   GAL   & ... & 262.07 & 4.52 & 0.29 & $<1.08$ & --10.55 \\
1730--130   &   J1733.0--1308    &   LPQ   & ... & 263.27 & --13.14 & 0.90 & 3.65 & --10.37 \\
S4 1739+52   &   J1740.0+5209    &   HPQ   &   C   & 265.02 & 52.16 & 1.38 & 1.11 & --9.96 \\
RGB J1742+597   &   J1742.1+5947    &   BLO   & ... & 265.54 & 59.80 & 0.40 & ... & --10.93 \\
NPM1G +19.0510   &   J1744.2+1934    &   BLO   & ... & 266.07 & 19.57 & 0.08 & $<0.2$ & --11.08 \\
1741--038   &   J1744.6--0354    &   HPQ   &   C   & 266.15 & --3.91 & 1.05 & 2.44 & --10.76 \\
S4 1749+70   &   J1748.5+7004    &   BLO   & ... & 267.13 & 70.08 & 0.77 & ... & --10.70 \\
B3 1747+433   &   J1749.0+4323    &   BLO   & ... & 267.27 & 43.39 & ... & ... & ... \\
PKS 1749+096   &   J1751.5+0937    &   BLO   &   C   & 267.89 & 9.63 & 0.32 & 6.28 & --10.10 \\
RX J1756.2+5522   &   J1756.6+5524    &   BLO   & ... & 269.15 & 55.40 & ... & ... & ... \\
S5 1803+784   &   J1800.4+7827    &   BLO   &   C   & 270.12 & 78.47 & 0.68 & 1.94 & --10.36 \\
3C 371.0   &   J1807.0+6945    &   BLO   &   C   & 271.75 & 69.76 & 0.05 & 1.21 & --10.50 \\
B2 1811+31   &   J1813.4+3141    &   BLO   & ... & 273.36 & 31.69 & 0.12 & ... & --10.71 \\
4C 56.27   &   J1824.0+5651    &   BLO   &   C   & 276.01 & 56.87 & 0.66 & 1.54 & --10.41 \\
1828+487   &   J1829.8+4845    &   LPQ   &   C   & 277.47 & 48.75 & 0.69 & 2.31 & --11.12 \\
RX J1829.4+5403   &   J1829.8+5404    &   BLO   & ... & 277.47 & 54.08 & ... & ... & ... \\
RX J1838.7+4802   &   J1838.6+4756    &   BLO   & ... & 279.66 & 47.95 & ... & ... & ... \\
J1849+6705   &   J1849.3+6705    &   GAL   &   C   & 282.32 & 67.09 & 0.66 & 3.50 & --9.78 \\
1851+488   &   J1852.5+4853    &   QSO   & ... & 283.14 & 48.90 & 1.25 & 0.45 & --10.60 \\
RX J1903.1+5540   &   J1903.0+5539    &   BLO   & ... & 285.77 & 55.65 & ... & ... & ... \\
RX J1931.1+0937   &   J1931.2+0939    &   BLO   & ... & 292.81 & 9.65 & ... & ... & ... \\
1ES 1959+650   &   J2000.0+6508    &   BLO   & ... & 300.02 & 65.13 & 0.05 & $<0.5$ & --10.10 \\
S5 2007+77   &   J2006.0+7751    &   BLO   &   C   & 301.51 & 77.86 & 0.34 & $<0.91$ & --10.78 \\
S5 2010+72   &   J2009.1+7228    &   BLO   & ... & 302.29 & 72.47 & ... & 0.44 & ... \\
PKS 2012--017   &   J2015.3--0129    &   BLO   & ... & 303.83 & --1.49 & 0.52 & $<0.2$ & --10.73 \\
J2017+0603   &   J2017.3+0603    & ... & ... & 304.34 & 6.06 & ... & $<0.16$ & ... \\
PKS 2022--077   &   J2025.6--0735    &   QSO   & ... & 306.42 & --7.60 & 1.39 & 1.18 & --9.68 \\
PKS 2032+107   &   J2035.4+1100    &   BLO   & ... & 308.86 & 11.00 & 0.60 & 0.62 & --10.41 \\
J2049+1003   &   J2049.7+1003    & ... & ... & 312.44 & 10.05 & ... & 0.39 & ... \\
PKS 2047+039   &   J2050.1+0407    &   BLO   & ... & 312.54 & 4.13 & ... & 0.38 & ... \\
RBS 1752    &   J2131.7--0914    &   BLO   & ... & 322.93 & --9.24 & 0.45 & ... & --11.00 \\
2131--021   &   J2134.0--0203    &   HPQ   &   C   & 323.51 & --2.06 & 1.29 & 1.83 & --11.08 \\
2141+175   &   J2143.4+1742    &   QSO   & ... & 325.87 & 17.72 & 0.21 & 0.75 & --10.08 \\
2144+092   &   J2147.2+0929    &   QSO   & ... & 326.82 & 9.50 & 1.11 & 0.94 & --9.96 \\
2145+067   &   J2148.5+0654    &   LPQ   &   C   & 327.13 & 6.90 & 0.99 & 4.72 & --10.81 \\
PKS 2149+17   &   J2152.5+1734    &   BLO   & ... & 328.14 & 17.58 & ... & 0.43 & ... \\
2155+312   &   J2157.4+3129    &   QSO   & ... & 329.37 & 31.50 & 1.49 & 0.74 & --10.36 \\
BL Lac   &   J2202.8+4216    &   BLO   &   C   & 330.72 & 42.28 & 0.07 & 2.91 & --10.09 \\
2201+171   &   J2203.5+1726    &   QSO   &   C   & 330.88 & 17.44 & 1.08 & 1.03 & --10.16 \\
PKS 2209+236   &   J2212.1+2358    &   QSO   & ... & 333.03 & 23.97 & 1.12 & 0.71 & --10.88 \\
3C 446   &   J2225.8--0457    &   BLO   &   C   & 336.47 & --4.96 & 1.40 & 6.89 & --10.31 \\
2227--088   &   J2229.7--0832    &   HPQ   &   C   & 337.44 & --8.53 & 1.56 & 2.77 & --9.83 \\
2230+114   &   J2232.5+1144    &   HPQ   &   C   & 338.13 & 11.74 & 1.04 & 4.71 & --10.04 \\
2234+282   &   J2236.2+2828    &   HPQ   &   C   & 339.06 & 28.47 & 0.80 & 1.26 & --10.21 \\
RGB J2243+203   &   J2244.0+2021    &   BLO   & ... & 341.01 & 20.36 & ... & ... & ... \\
B3 2247+381   &   J2250.1+3825    &   BLO   & ... & 342.53 & 38.43 & 0.12 & $<0.32$ & --10.76 \\
3C 454.3   &   J2253.9+1608    &   HPQ   &   C   & 343.49 & 16.15 & 0.86 & 17.32 & --9.09 \\
BZB J2304+3705  &   J2304.3+3709    &   BLO   & ... & 346.10 & 37.16 & ... & ... & ... \\
TXS 2320+343   &   J2322.6+3435    &   BLO   & ... & 350.65 & 34.58 & 0.10 & ... & --11.03 \\
PKS 2320--035   &   J2323.5--0315    &   QSO   & ... & 350.89 & --3.26 & 1.41 & 1.06 & --10.43 \\
1ES 2321+419   &   J2323.5+4211    &   BLO   & ... & 350.89 & 42.19 & 0.06 & $<0.4$ & --10.53 \\
B3 2322+396   &   J2325.2+3957    &   BLO   & ... & 351.32 & 39.96 & ... & 0.35 & ... \\
PKS 2325+093   &   J2327.7+0943    &   QSO   & ... & 351.93 & 9.73 & 1.84 & 2.57 & --9.85 \\
RX J2338.9+2124  &   J2339.0+2123    &   BLO   & ... & 354.77 & 21.39 & 0.29 & ... & --11.02 \\
1ES 2344+514   &   J2347.1+5142    &   BLO   & ... & 356.78 & 51.71 & 0.04 & $<0.38$ & --10.67 \\
\end{longtable}
\tablefoot{
Columns 7, 8, and 9 list the redshift, 37 GHz flux density,  and \fermilat energy flux, respectively. The fluxes are averaged over the 1FGL period. The redshifts were collected from Simbad (http://simbad.u-strasbg.fr/simbad/sim-fid) and NED (http://nedwww.ipac.caltech.edu/) databases.
}
}

\end{document}